\documentstyle[11pt]{article}
\newcommand{\be}{\begin{equation}}
\newcommand{\ee}{\end{equation}}
\newcommand{\ba}{\begin{eqnarray}}
\newcommand{\ea}{\end{eqnarray}}
\newcommand{\bann}{\begin{eqnarray*}}
\newcommand{\eann}{\end{eqnarray*}}
\newcommand{\ul}{\underline}

\begin{document}
\hbadness=10000
\setcounter{page}{1}

\title{{\bf \huge  Thermal photon production\\  in heavy ion collisions}}
%\vspace{-3.0cm}
%\hspace{-1.0cm}
%\hspace*{\fill}
%{\normalsize DMR-THEP-94-8/W} \\*[-3.5ex]
%\hspace*{\fill}
%{\normalsize April 1994} \\*[1.0ex]
{\huge \bf }

\author{N. Arbex$^1$\thanks{E. Mail: ARBEX@CONVEX.HRZ.UNI-MARBURG.DE},
U. Ornik$^2$\thanks{E.Mail: ORNIK@RZRI6F.GSI.DE}, M.
Pl\"umer$^1$\thanks{E. Mail: PLUEMER\_M@VAX.HRZ.UNI-MARBURG.DE},
A. Timmermann$^1$\thanks{E. Mail: TIMMERMANN@VAX.HRZ.UNI-MARBURG.DE}{ }
 and R.M.
Weiner$^1$\thanks{E. Mail: WEINER@VAX.HRZ.UNI-MARBURG.DE} }

\date{$^1$ Physics Department, Univ. of Marburg, Marburg, FRG\\
      $^2$ GSI, Darmstadt, FRG}

\maketitle

\vspace{0.5cm}

\begin{abstract}
Using a three-dimensional
hydrodynamic simulation of the collision and an
equation of state containing a first order phase transition
to the quark-gluon plasma,
we study thermal photon production for $Au+Au$ collisions
at $E_{lab}=11.5$ AGeV and for $Pb+Pb$ collisions at $160$ AGeV.
We obtain surprisingly high rates of thermal
photons even at the lower energy,
suggesting that, contrary to what was expected so far,
photon production may
be an interesting topic for experimental search also at the Alternating
Gradient Synchrotron.
When applied to the reaction $S+Au$ at $200$ AGeV, our model can
reproduce preliminary data obtained by the WA80 Collaboration without
having to postulate the existence of an extremely long-lived mixed
phase as was recently proposed.
%This result is a consequence of the high degree of stopping
%and the creation of a long-lived mixed phase.
%Our predictions for the rates at this energy are comparable to the
%experimentally observed photon rates for $S+Au$ collisions at $E_{lab}=
%200 $AGeV.

\end{abstract}

\newpage

In ultrarelativistic heavy-ion collisions one of the most
challenging goals is to study a possible phase transition between nuclear
matter and the quark-gluon-plasma (QGP),
consisting of deconfined quarks and gluons.
This expectation is supported by lattice gauge calculations
and phenomenological models.
Photons are a very promising probe in the
experimental search for the QGP which is
expected to exist for a brief period of a few fm/c before the
majority of the final state hadrons is emitted.
Particles which interact only electromagnetically may be regarded as
free particles, carrying information on the early stage of
the collision.
In order to describe the space-time development of a
nucleus-nucleus collision, very often hydrodynamical models are used,
which are based on the concept of local thermal equilibrium.
For heavy ion collisions at the Brookhaven
Alternating Gradient Synchrotron (AGS) and at the
CERN Super Proton Synchrotron (SPS)
this condition may be
fullfilled because of the high degree of stopping.
Thus,
high compression of nuclear matter has to be considered and energy densities
near the critical value for the deconfinement phase transition may
be reached.

In the present paper,
we study thermal photon production in a realistic hydrodynamical
framework and make predictions for measurements at AGS and SPS energies.
We focus on the reactions $Au+Au$ at $E_{lab}=11.5$ AGeV
and $Pb+Pb$ at $160$ AGeV.
We also apply our model to $S+Au$ collisions at $200$ AGeV
and compare our results to preliminary data on photon production
obtained by the WA80 Collaboration \cite{wa80}.

At the SPS, lead beams will be available at the end of 1994 and direct
photon production measurements will be performed by the WA98 Collaboration
\cite{wa98}.
Furthermore, it is intended \cite{tannenbaum}  to test
the PHENIX-detector \cite{nagamiya},
which is designed to detect photons with high accuracy and resolution,
at the AGS in 1996.
This means that
our predictions for $Pb+Pb$ at $160$ AGeV and for $Au+Au$ at
$11.5$ AGeV can be checked in the near future.

%To be specific,  we modify the classical Landau full stopping initial
%conditions by simulating the collision stage with the computer code
%HYLANDER (3-d Hydrodynamical Expansion Routine). We then estimate the
%production rates of thermal photons, taking into account
%the contribution of $A_1$-mesons in the mixed and hadronic phase.

% Previous work on photon production was limited to the
%CERN Super Proton Synchrotron (SPS) or
%higher energies because the number of photons expected at AGS energies was
%considered to be too low. Our results suggest that this opinion may
%have to be revised.

The photon production rates in thermalized matter have been calculated
both for a QGP and for a hadron gas.
In a QGP of temperature $T$, the rates for a baryochemical potential
$\mu=0$ can be expressed as \cite{kapus1}:
\begin{equation}
E_{\gamma} \frac{dN}{d^3kd^4x}
= \frac{5}{18 \pi^2} \alpha \alpha_s
T^2 e^{-E_\gamma/T} \ln \frac{0.2317 E_\gamma}{\alpha_s T}.
\label{eq:qrate}
\end{equation}
where $E_\gamma$ and $\vec{k}$ are the energy and three-momentum of
the photon, and where
$\alpha=1/137$ and $\alpha_s$ are the electromagnetic and
the temperature dependent strong coupling constant, respectively.
{}From lattice results $\alpha_s (T)$ can be parametrized as \cite{karsch}:
\begin{equation}
\alpha_s(T) = \frac{6 \pi}{(33-2n_f)\ln(8 T/T_c)}.
\label{eq:alphas}
\end{equation}
where $T_c$ is the critical temperature and
$n_f$ is the number of flavors (below, we take $n_f=2$).

For a hadronic resonance gas at temperature $T$ and baryochemical
potential $\mu=0$, the thermal photon rates were calculated in
\cite{shuryak}. The authors of \cite{shuryak} parametrized
their results in the form
\begin{equation}
E_{\gamma} \frac{dN}{d^3kd^4x} =2.4\ T^{2.15}\ \exp\left[\frac{-1}
{(1.35 T E_{\gamma})^{0.77}}
-\frac{E_{\gamma}}{T}\right]\  \qquad [GeV^{-2} fm^{-4}],
\label{eq:hgrate}
\end{equation}
where $E_{\gamma}$ and $T$ are measured in units of GeV.
Eq. (\ref{eq:hgrate}) contains the contributions
{}from $A_1$-resonances which were shown to be important in \cite{shuryak}
and which had not been taken into account in earlier calculations
of the rates \cite{nadeau}.

%The dependence of the rates
%on the baryochemical potential is expected to be small for the hadron gas
%as processes involving baryons do not contribute strongly to photon
%production.
In the mixed phase the rates read
\begin{equation}
\left. E_{\gamma} \frac{dN}{d^3kd^4x} \right|_{mix} =
w(\varepsilon) \left. E_{\gamma} \frac{dN}{d^3kd^4x}\right|_{QGP}
+ (1-w(\varepsilon)) \left. E_{\gamma} \frac{dN}{d^3kd^4x}\right|_{had}.
\end{equation}
where $w(\varepsilon)$ is the fraction
of QGP at the energy density $\varepsilon$.

In order to obtain the single-inclusive spectra it is necessary
to integrate the photon rates over the space-time region defined by
the space-time evolution of the hot and dense matter,
\begin{equation}
E_{\gamma} \frac{dN}{d^3k}  = \int d^4x E_{\gamma} \frac{dN
}{d^3kd^4x}  \left(T(x),u^{\mu}(x)\right)\qquad ,
\end{equation}
where the temperature field $T(x)$ and the four velocity field
$u^\mu(x)$ are obtained from a hydrodynamic description of the
expanding matter. On the rhs of eqs. (\ref{eq:qrate}) and (\ref{eq:hgrate})
the energy $E_{\gamma}$ has to be replaced by $k_\mu u^\mu (x)$.
In the framework of hydrodynamics the ultrarelativistic
heavy ion collision can be divided into three stages:
(i) the collision or compression stage, (ii) the expansion stage and
(iii) the freeze-out stage.
The space-time development of dense and hot matter in local thermal
equilibrium is
described by the equations  of relativistic hydrodynamics:
\be
\partial_{\mu} T^{\mu \nu} = 0 , \qquad \partial_{\mu} (b u^{\mu}) = 0,
\ee
where $T^{\mu \nu}$ is the energy-momentum tensor,
$b$ the baryon-number density and $u^{\mu}$ the four-velocity of the
fluid element. Here we consider the case of an ideal fluid, i.e.,
we neglect dissipative effects. In this case,
$T^{\mu\nu} = (\varepsilon +P) u^\mu u^\nu -g^{\mu\nu} P$,
where $\varepsilon$ is the energy density $P$ the pressure.
In order to solve the  equations, one needs to specify the
equation of state (EOS), which may be given, e.g., in the form
$P=P(\varepsilon ,b)$. The equations of motion
are solved exactly and fully 3-dimensionally \cite{udo}
with the numerical code HYLANDER.

Our hydrodynamical simulation describes also the initial
compression (or collision) stage
where the propagation of shock waves which heat up the system is
calculated fully 3-dimensionally.
The applicability of one-fluid hydrodynamics at this early stage is supported
by the fact that in heavy ion collisions at AGS and at SPS energies the
mean free
paths are small compared to the size and lifetime
of the system
and a considerable amount of stopping is expected\footnote{RQMD
simulations for central $Au+Au$
collisions have found \cite{RQMD}
complete stopping up to (and even somewhat beyond) SPS energies.}.

We use an EOS given by a parametrization \cite{udo}
of lattice-QCD results \cite{lattice} where the hadronic
phase is treated as a resonance gas.
This EOS exhibits a first order
phase transition between hadronic matter and the QGP at a critical
temperature\footnote{The results in \cite{lattice} were obtained for a
purely gluonic system. New results which take into account the effects
of dynamical quarks yield lower values for the critical temperature,
$T_c \sim 150\ MeV$ (see \cite{newlat} and refs. therein). There is no EOS
published yet for this case.}
$T_c=200\ MeV$; it corresponds to a baryochemical
potential $\mu=0$ \footnote{At present there exist no reliable
lattice data concerning the EOS
at $\mu \not= 0$.}. Apart from the EOS our hydrodynamic model does not
contain any free parameters.

Fig. 1 shows the first $8$ fm/c of the space-time evolution of
central $Au+Au$ collisions at
$E_{lab} = 11.5$ AGeV (left column) and of central $Pb+Pb$ collisions
at $160$ AGeV, in time steps $\Delta t=2$ fm/c.
Contour plots for the energy-density calculated with HYLANDER are displayed.
Four characteristic regimes are considered.
The energy density $\varepsilon=0.14$ GeV/fm$^3$ corresponds to the
interface between hot and cold nuclear matter, $\varepsilon=0.25$ GeV/fm$^3$
to the freeze out regime, the region $2.5 \le \varepsilon \le 5.5$ GeV/fm$^3$
to the mixed phase and $\varepsilon \ge 5.5$ GeV/fm$^3$
to the pure QGP phase, respectively. As can be seen in the figure, for
both reactions  a lump of pure QGP and a mixed phase
with a lifetime of about $7$ fm/c are produced.
The total lifetimes of the thermalized matter are
$\sim 12$ fm/c for $Au+Au$
at $11.5$ AGeV and $\sim 16$ fm/c for $Pb+Pb$ at $160$ AGeV.
Figs. 2a and 2b show the thermal photon spectra for these two reactions.
The rates are surprisingly large even at AGS energies.
To understand the origin of these results,
we have plotted separately the contributions from the pure QGP phase,
the mixed phase and the purely hadronic phase.
For $Au+Au$ at $11.6$ AGeV the
contribution of the purely hadronic phase dominates
the photon production, whereas for $Pb+Pb$ at $160$ AGeV it is comparable
to the contribution of the pure QGP phase.
The reason that the pure hadronic phase plays such an important role
in photon production is twofold.
Firstly, as can be
seen from  eqs. (\ref{eq:qrate}) and (\ref{eq:hgrate}),
at temperatures $T\sim 0.2$ GeV the hadron gas outshines the QGP
by a factor of about two.
Secondly, the hadronic space-time
volume exceeds the volumes occupied by the mixed phase
and by the QGP.
The fact that at SPS energies the QGP contribution can compete with that
of pure hadronic phase is a consequence of the high initial
temperatures ($T_i \sim 300$ MeV) at these energies.

Preliminary data on photon production for $S+Au$ collisions at SPS energies
were recently presented by the WA80 Collaboration \cite{wa80}.
The fact that the
measured direct photon rates are surprisingly high
has led to speculations concerning the existence
of an extremely long-lived mixed phase
with lifetimes of about $30-40$ fm/c\cite{shuryak1}. We have applied
our hydrodynamic collision simulation to this reaction\footnote{For the
system $S+Au$ one expects a smaller degree of stopping than for $Pb+Pb$
or $Au+Au$. On the other
hand, fits to rapidity and transverse momentum spectra of hadrons
produced in $S+S$ at $200$ AGeV already indicate a considerable amount
of stopping \cite{jan}
which should be even more pronounced for $S+Au$. This suggests
that our hydrodynamic collision simulation is also
applicable for the latter reaction.
Note, however, that recently the WA80 data have also been described
\cite{seibert,sinha} under the assumption of Bjorken initial conditions.}
and plotted
the resultant thermal photon rates in Fig. 2c.
We find a remarkably good agreement
with the WA80 data \cite{wa80}, without having to adjust any parameters or
having
to assume an extremely long-lived mixed phase as in Ref. \cite{shuryak1}.
In particular, it turns out that the hadronic rather than the mixed phase
dominates photon emission.
The space-time evolution for $S+Au$ collisions is illustrated in Fig.3.
Due to the asymetry of the collision, the central rapidity region
is shifted, which is of importance for the calculation of the photon spectra.

%Deviations in the $k_{\perp}$-range of $\approx 1$ GeV
%could be due to effects of nonvanishing $\mu$ in the mixed and
%hadronic phase or to an incomplete extraction of photons from
%resonance decays in this region. Also hard bremsstrahlung
%processes may contribute in this region.

So far we have not discussed the effects of baryon stopping on thermal
photon production. At AGS and SPS energies, one expects nonvanishing
values of the baryochemical potential due to the high baryon number
densities in the central rapidity region.
In Ref. \cite{traxler},
photon emission rates from a QGP of temperature $T$ and baryochemical
potential $\mu$ were calculated by means of the Braaten-Pisarski technique.
The authors give the expression
\begin{equation}
E_{\gamma} \frac{dN}{d^3kd^4x}
= \frac{5}{18 \pi^2} \alpha \alpha_s
T^2(1+\frac{\mu^2}{\pi^2 T^2}) e^{-E_\gamma/T} \ln
\frac{0.2317 E_\gamma}{\alpha_s T}.
\label{eq:murate}
\end{equation}
At fixed baryon number density and energy density, we have used a bag model
EOS with $T_c(\mu=0)=0.2$ GeV to estimate the decrease in temperature and the
resultant decrease in the photon rates  \cite{dumitru,traxler} near the
critical curve of the deconfinement phase transition. Using
eqs. (\ref{eq:murate}) and (\ref{eq:hgrate}) we obtain reduction factors
of about $2$ both for the QGP and for the hadronic
component\footnote{
A reduction by a factor of $\sim 2$ for $S+Au$ at $200$ AGeV
would imply that we underestimate the
WA80 data. There are three possible effects which would lead to an
increase of the photon rates and thus may explain the
missing factor.
An additional contribution is also expected from the
hadronic processes $A_1 \rightarrow \pi \gamma$,
$b_1 \rightarrow \pi\pi^0 \gamma$,
and $K_1 \rightarrow K \gamma$ \cite{haglin}.
%To begin with, the strong coupling constant $\alpha_s$
%could have larger values than those given by the expression
%(\ref{eq:alphas}).
Furthermore at nonzero baryochemical potential the presence of
baryons may open additional channels for thermal photon
production. Thirdly, one expects a decrease
of the rho meson mass as the temperature approaches the critical temperature
of the chiral phase transition\cite{chiral}.}.

{}From the above considerations we conclude that the main result
of this letter -- namely, that there are experimentally observable
rates of thermal photons to be expected both at the AGS and at the SPS  --
remains unaffected if one takes into account a finite baryochemical potential.
\\[3ex]

%As was mentioned in the introduction, direct photon measurements will
%be performed for experiments scheduled at the SPS
%in 1994/95 and at the AGS in 1996, which implies that our prediction
%can be checked in the near future.\\[3ex]

%\newpage

This work was supported by the Federal Minister of Research and Technology
under contract 06MR731 and the Gesellschaft f\"ur Schwerionenforschung.
 N. Arbex acknowledges a CNPq fellowship.
We would like to thank P. Rehberg and M.H. Thoma for interesting
discussions.

\newpage

{\Large \bf Figure Captions}\\
\begin{description}
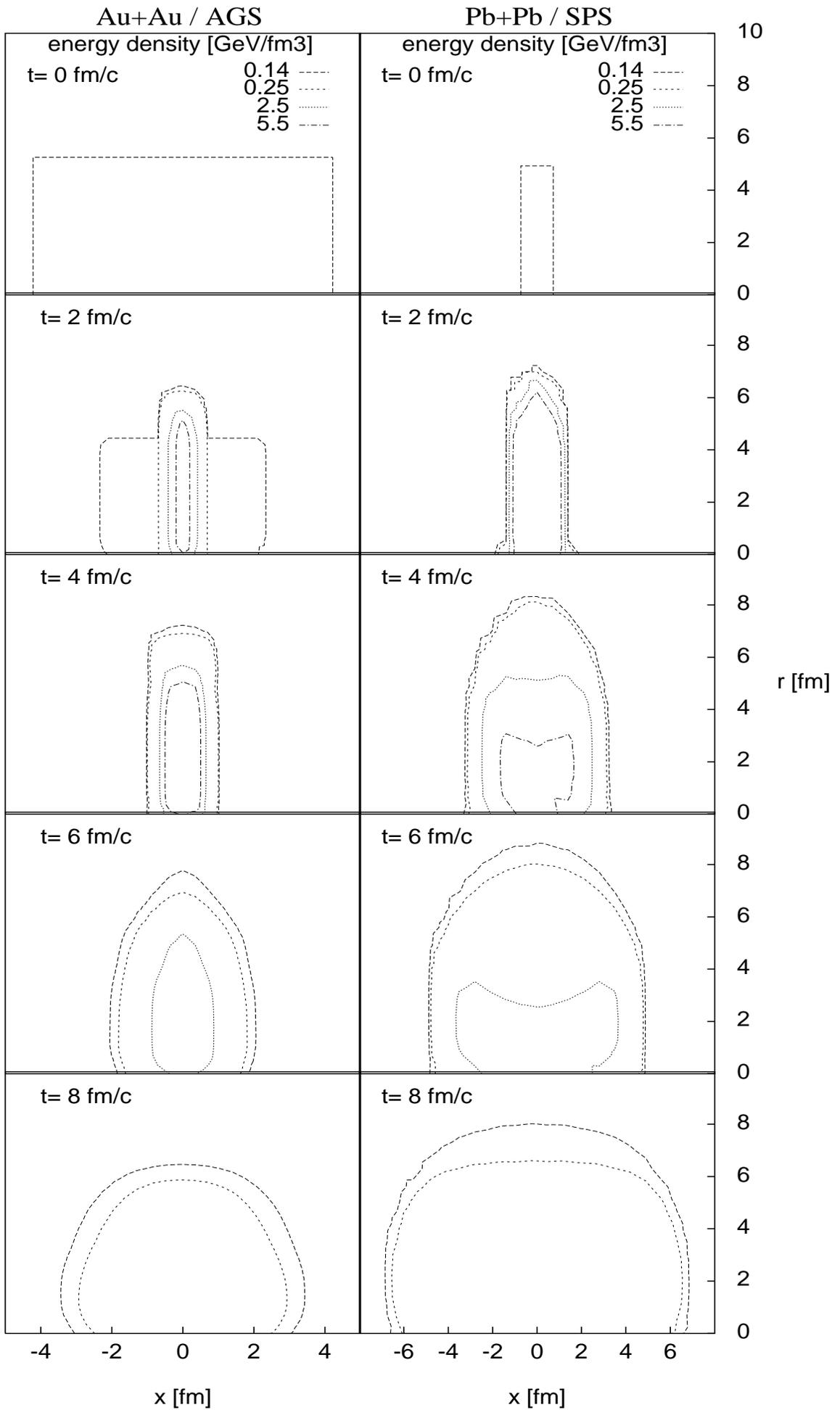
\item[Fig. 1]  Energy density contour plots in the $(x,r)$ plane,
for $Au+Au$ collisions
at $E_{lab}=11.5$ AGeV (left column)  and $Pb+Pb$ collisions at
$E_{lab}=160$ AGeV (right column). $x$ is the coordinate
in longitudinal direction, $r$ the radial coordinate in the transverse
plane. The plots were obtained in the equal velocity frame by applying
the hydrodynamic collision simulation.
\item[Fig. 2] Single-inclusive photon
spectra as a function of
the transverse momentum $k_{\perp}$ for mid-rapidity photons,
(a) for $Au+Au$ collisions
at $E_{lab}=11.5$ AGeV, (b) for $Pb+Pb$ at $160$ AGeV  and (c) for
$S+Au$ at
$200$ AGeV. The contributions from the pure QGP phase, the mixed phase
and the pure hadronic phase
are shown separately. The data points are from ref. \cite{wa80}.
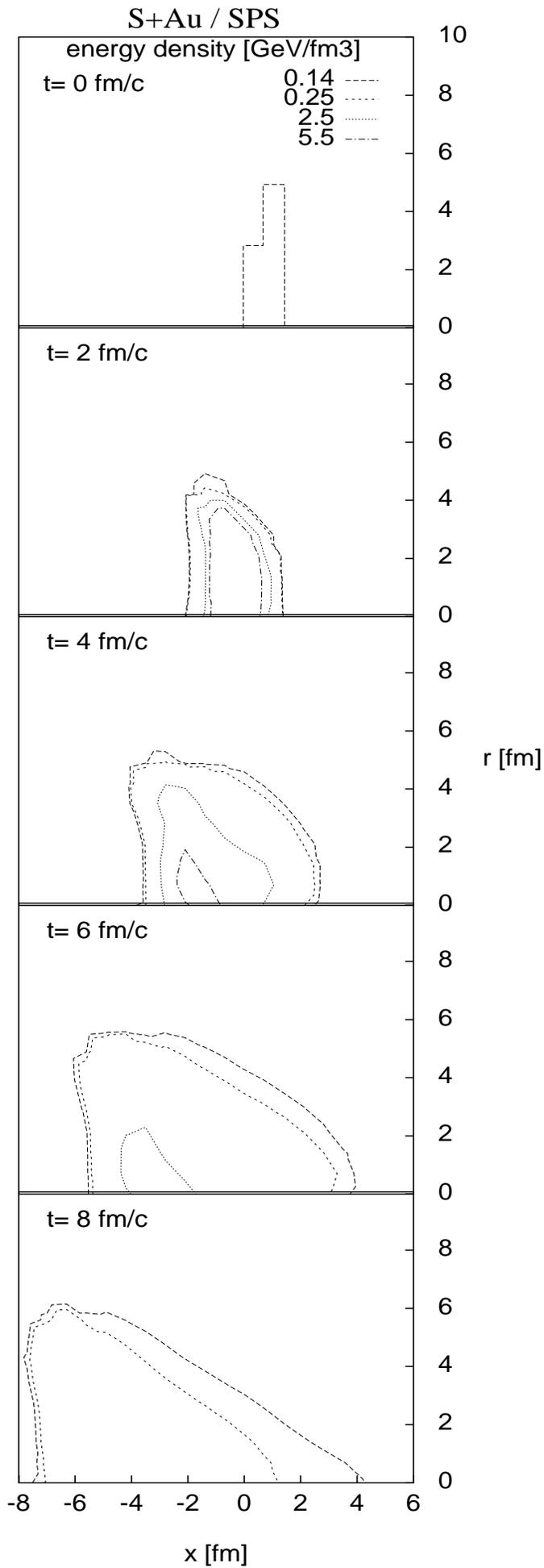
\item[Fig. 3] Energy density contour plots as in Fig.1
for $S+Au$ collisions at $E_{lab}=200$ AGeV.
\end{description}

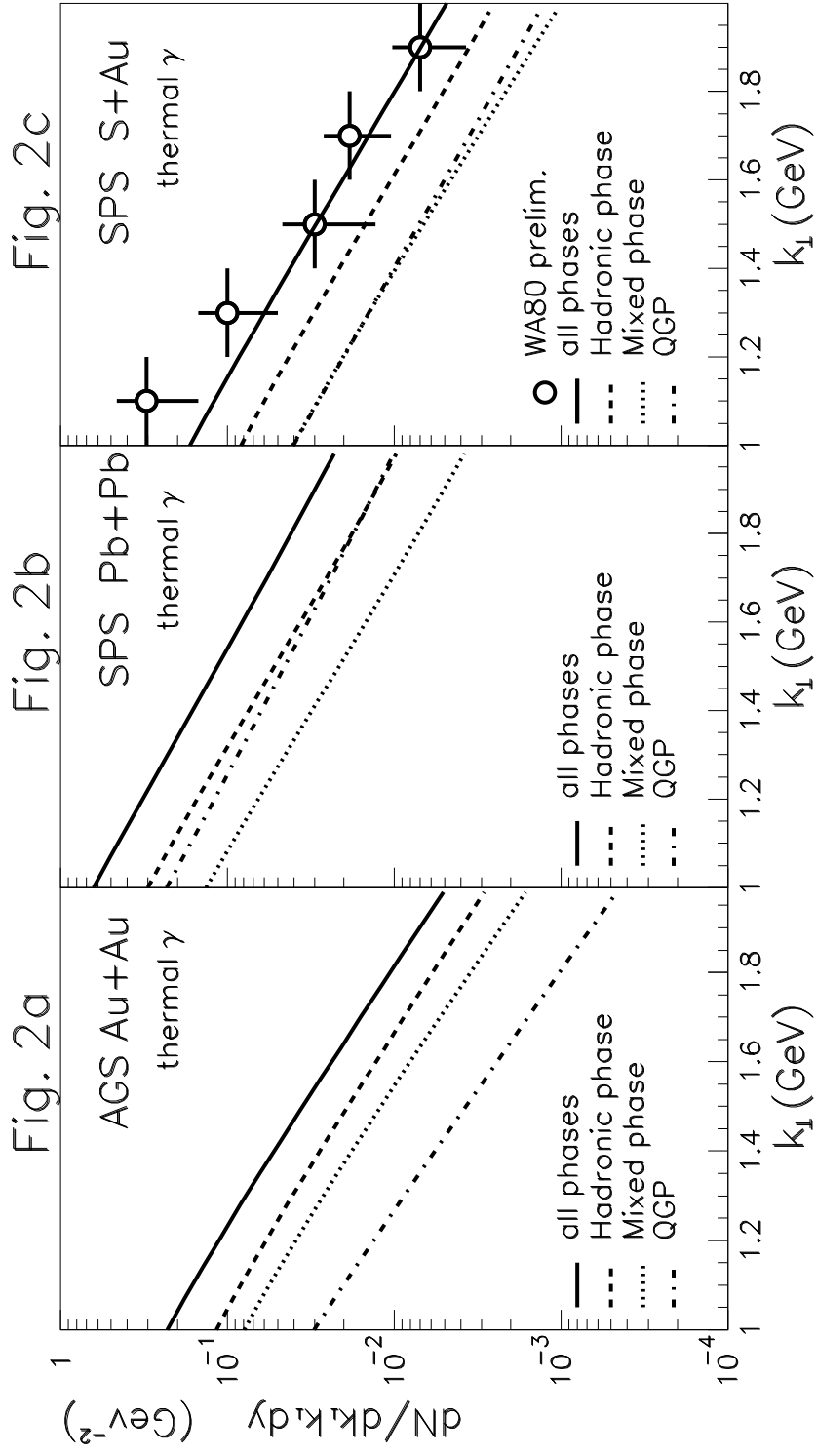
\end{document}